\documentstyle[preprint,aps]{revtex}
\begin{document}
\draft

\title{Correcting the effects of spontaneous emission \\on cold-trapped ions}
\author{C. D'Helon and G.J. Milburn}
\address{Department of Physics University of Queensland St Lucia 4072 Australia 
\\ {\em email: dhelon@wilson.physics.uq.oz.au}}
\maketitle

\begin{abstract}

We propose two quantum error correction schemes which increase the maximum 
storage time for qubits in a system of cold trapped ions, using a minimal number of 
ancillary qubits.
Both schemes consider only the errors introduced by the decoherence due to 
spontaneous emission from the upper levels of the ions.
Continuous monitoring of the ion fluorescence is used in conjunction with selective 
coherent feedback to eliminate these errors immediately following spontaneous 
emission events, 
{ and the conditional time evolution between quantum jumps 
is removed by symmetrizing the quantum codewords.}

\end{abstract}
\newpage

\section{Introduction}
\label{Introduction}

It was recognized over a decade ago by Deutsch \cite{Deutsch85} that a quantum computer 
has the potential to perform certain computational tasks much more efficiently than its 
classical counterparts, and the most striking example to date is the factorization of 
large numbers proposed by Shor \cite{Shor94}.
The advent of quantum computation would also enable the realization of arbitrary quantum 
measurement processes and ultimately the efficient simulation of physical systems as 
envisaged by Feynman \cite{Feynman}.
A measurement scheme recently presented by the authors \cite{Dhelon96} illustrates the 
usefulness of quantum computations in generating a ``difficult'' operator transformation in 
the context of an ion trap.

The possibility of building a quantum computer has received a lot of attention in the last few 
years, following several implementation proposals based on well-known physical systems.
At this stage, ion traps \cite{Cirac95} and optical cavities \cite{Turchette95} are the leading 
candidates, and experimental work has progressed to the point of demonstrating the operation 
of two-bit quantum gates \cite{Monroe95}.

A major hurdle in the implementation of a quantum computer is that the presence of decoherence 
in any open physical system introduces random errors in the computation process, which will grow 
with time if left unchecked.

To illustrate the importance of this problem, Plenio and Knight \cite{Plenio96} have considered 
the fundamental limitation imposed on Shor's factorization algorithm by decoherence due to 
spontaneous emission.
The maximum computation time $T$ available on an $N$-qubit register is bounded by the 
decoherence time $\tau_d$, which can be expressed as
\begin{equation}
\tau_d=\frac{\tau_{q}}{N}
\hspace{2.5 mm},
\end{equation}
if the qubits are coupled to the environment independently, where 
$\tau_{q}$ is the decoherence time of a single qubit.
In the case of a linear rf trap, the switching rate of the laser pulses used to implement quantum 
gates is directly proportional to the Rabi frequency $\Omega$, which depends on the 
spontaneous emission rate $\gamma$ of the qubit transitions 
($\Omega\propto\sqrt{\gamma}$), 
therefore the execution time of an elementary logic gate ultimately depends on the qubit 
decoherence time $\tau_{q}=\gamma^{-1}$.
Thus the dependence of the computation time on the number of input bits ($\log_2L$ bits for 
a decimal integer $L$) was shown to be much stronger
\begin{equation}
T\propto(\log_2L)^8
\hspace{2.5 mm},
\end{equation}
than that obtained by assuming the logic gates can be performed in a time independent of 
the qubit decoherence time,
\begin{equation}
T\propto(\log_2L)^3
\hspace{2.5 mm}.
\end{equation}
This result implies that factorizing a 4-bit number using Shor's algorithm would be very difficult 
to implement experimentally, whereas the factorization of nontrivial numbers with hundreds of 
bits appears impossible for the trapped-ion realization of a quantum computer.

Hence quantum error-correcting codes are necessary to run error-free computations, and a number
of recent papers \cite{Shor95,Laflamme96,Steane96,Calderbank96} have addressed this issue.
The proposed codes allow for an arbitrary interaction between qubits and their environments, and 
prevent memory errors by redundantly encoding the information contained in logical qubits across 
an entanglement of several physical qubits.
The resulting entangled states are known as {\bf quantum codewords}, which can be decoded to 
give logical qubits, and the additional qubits required to implement the codes are labeled the 
{\bf ancilla}.

The codewords can be recovered after each decoherence event using coherent feedback, given 
that only one qubit out of each codeword decoheres.
Laflamme {\em et al.}\ \cite{Laflamme96} have shown that the most efficient perfect code 
which can correct all one-bit errors using a minimum number of ancillary qubits, consists of 
five-qubit codewords.

The problem of additional errors introduced by decoherence and inaccuracies during the 
error correction process was recently addressed by DiVincenzo and Shor \cite{DiVincenzo96}.
The quantum gates used to implement error-correcting codes often depend on analog 
parameters, hence they need to be fault-tolerant in order to achieve robust computation 
which yields correct results even in the presence of moderate levels of noise.

In contrast to the general schemes referred to above, we only consider the errors arising 
from one source of decoherence, namely spontaneous emission, in a quantum computer 
based on an ion trap.
This implies that our error-correcting codes are limited in scope, however they are directly 
applicable to experiment, and use a minimum of additional memory resources.

Our physical system consists of N ions confined in a linear rf trap, as proposed by Cirac and 
Zoller \cite{Cirac95} for the implementation of a quantum computer.
Each of the ions is laser-cooled into the Lamb-Dicke limit, and experiences harmonic motion 
at the trapping frequency $\nu$ of a collective vibrational mode.
Qubits are represented by two-level electronic transitions $|0\rangle\leftrightarrow|1\rangle$ 
consisting of hyperfine levels in order to minimize the atomic recoil from spontaneous 
emission events.
One-bit and two-bit quantum gates are realized by applying carrier and sideband Raman laser 
pulses to the appropriate ion(s), using the collective vibrational mode as a common bus to 
achieve ion-ion coupling.

The errors experienced by an N-ion register running a quantum computation can be conveniently 
divided into memory and gate errors.
Memory errors affect qubits when they are simply stored in the trap, and are due to decoherence 
from a number of sources including spontaneous emission from qubits, vibrational heating due 
to technical imperfections in the trap, and collisions with residual gas molecules.
Gate errors arise during quantum gate operations due to decoherence from the same sources as 
above, or from technical inaccuracies involved in applying a gate, such as the timing of laser 
pulses.

A quantum error correction scheme rendering quantum gates tolerant to errors introduced by 
vibrational damping of the ions was recently proposed by Cirac {\em et al.}\ \cite{Cirac96}.
Their scheme redundantly encodes logical qubits in four electronic levels of each ion, and 
inverts the effects of quantum jumps of the phonon number using projection measurements 
to provide non-unitary feedback and restore the initial state, so that the gate can be repeated.
Fault-tolerance is achieved by applying a measurement projecting onto the ideal state, if no
jump occurs during the gate operation.
The size of the quantum network which can be realized without any errors is squared relative
to the uncorrected case, since the first order effects of decoherence are eliminated.

In this article we present two quantum error-correcting codes which address only those 
memory errors introduced by spontaneous emission from the upper levels of qubits 
$|1\rangle_j$ to the respective ground levels $|0\rangle_j$, at the decay rate $\gamma_j$ 
for the $j$-th ion.
Without any error correction, the maximum storage time for qubits in an N-ion register can 
be estimated by the decoherence time $\tau_d=\gamma^{-1}/$N, assuming that the 
spontaneous emission rates of the ions are equal but independent of each other.
The upper levels are typically chosen to be metastable levels, hence the decoherence time 
for a single qubit ($\tau_q=\gamma^{-1}$) can be as long as one minute.

In both of our schemes, the initial state of the ion register is prepared in an entangled state, 
ie.\ a quantum codeword $|\psi\rangle_i$ which belongs to a logical Hilbert subspace ${\cal H}_l$.
Quantum jumps corresponding to spontaneous emission events are continuously monitored using 
photodetectors positioned around the ion trap, and when a jump is detected, its effect on the 
system is immediately inverted using selective coherent feedback.
Mabuchi and Zoller \cite{Mabuchi96} have shown that the inversion of quantum jumps is 
possible if quantum codewords are uniquely mapped by the jumps into error states which can be 
transformed unitarily back to the logical subspace ${\cal H}_l$.
We assume that the spontaneous emission events from different ions are distinguishable, 
so that the feedback process can be applied selectively to the system to recover the initial 
codeword $|\psi\rangle_i$, in a time much shorter than the decoherence time of the ion 
register, thus ensuring that no spontaneous emission occurs before the codeword is restored.


{
As emphasized by Plenio {\em et al.}\ \cite{Plenio97} recently, the 
quantum codewords used for the register need to be invariant under the 
conditional time evolution between jumps, so that the state prior to each 
decay event is known to be a codeword.
The conditional time evolution causes the ion register to evolve non-unitarily 
according to the transformation
\begin{equation}
U_{c}(t)=\exp\left(-\frac{\Gamma t}{2}\sum_j |1\rangle_j\langle1|\right)
\hspace{2.5 mm},
\label{cond_evolution}
\end{equation}
where $\Gamma$ is the spontaneous emission rate of all the excited levels, 
and the state of the register has been transformed to an interaction picture 
which eliminates the free evolution of the ions.
Thus the quantum codewords have to be restricted to a logical Hilbert subspace 
consisting of register states which have the same number of excited ions.
}


In the following section we outline the use of an alternative logical basis consisting of 
Fourier-transformed states, and present an error correction scheme for individual 
qubits based on entangling two ions for each codeword.
We then consider an error correction scheme for an N-ion register, which uses 
complementary electronic number states to realize codewords in a very efficient manner.

\section{Error correction using Fourier-transformed states}

For the purposes of quantum computation with an ion trap, a Fourier-transformed basis $\tilde{L}$ 
can be used as an alternative to the logical basis $L$ comprising the electronic levels 
$|0\rangle, |1\rangle$.
The Fourier-transformed states $|\tilde{0}\rangle, |\tilde{1}\rangle$ making up $\tilde{L}$, 
are orthogonal superpositions of the electronic levels,
\begin{eqnarray}
|\tilde{0}\rangle_i & = & \frac{1}{\sqrt{2}}(|0\rangle_i+|1\rangle_i) \\
|\tilde{1}\rangle_i & = & \frac{1}{\sqrt{2}}(|0\rangle_i-|1\rangle_i)
\hspace{2.5 mm}.
\end{eqnarray}
The basis $\tilde{L}$ holds the advantage that both amplitude coefficients of an arbitrary qubit 
can be preserved when there is a spontaneous emission from the excited level $|1\rangle$, hence 
enabling us to recover the initial qubit.

One-bit quantum gates are implemented in basis $L$ by applying a standing wave $k$-pulse 
$V_i^k(\phi)$ (with laser phase $\phi$) to the $i$-th ion,
\begin{equation}
V_i^k(\phi)=\exp\left[\frac{-ik}{2}\left(|1\rangle_i\langle0|e^{-i\phi}+|0\rangle_i\langle1|e^{i\phi}\right)\right]
\hspace{2.5 mm},
\end{equation}
to rotate the respective electronic levels.
The laser pulse is on resonance with the electronic transition 
$|0\rangle_i\leftrightarrow|1\rangle_i$, 
and the equilibrium position of the ion is placed at an antinode of the standing wave for the 
duration of the pulse.
The same unitary transformation $V_i^k(\phi)$ also implements rotations in $\tilde{L}$.
For example, a laser pulse $V_i^{\pi/2}(-\pi/2)$ can be used to swap between the basis states of 
$L$ and $\tilde{L}$,
\begin{eqnarray}
|0\rangle_i & \rightarrow & |\tilde{0}\rangle_i
\label{pi/2-pulse1} \\
|1\rangle_i & \rightarrow &-|\tilde{1}\rangle_i  
\label{pi/2-pulse2}
\hspace{2.5 mm}.
\end{eqnarray}

Cirac and Zoller \cite{Cirac95} have shown how one-bit rotations and a series of sideband laser 
pulses, with the equilibrium positions of the ions at the nodes of the respective standing waves, 
can be used to implement a controlled-not gate $U_{ij}$ in basis $L$, for two ions $i,j$,
\begin{eqnarray}
U_{ij} |0\rangle_i |0\rangle_j &= & |0\rangle_i |0\rangle_j \nonumber \\
U_{ij} |0\rangle_i |1\rangle_j &= & |0\rangle_i |1\rangle_j \nonumber \\
U_{ij} |1\rangle_i |0\rangle_j &= & |1\rangle_i |1\rangle_j \nonumber \\
U_{ij} |1\rangle_i |1\rangle_j &= & |1\rangle_i |0\rangle_j
\hspace{2.5 mm},
\end{eqnarray}
employing a vibrational mode common to the ions.
The control qubit $i$ remains unchanged, whereas the target qubit $j$ flips in the case that the 
control qubit is set to $|1\rangle_i$.
An experimental realization of this quantum gate has been demonstrated recently by 
Monroe {\em et al.}\ \cite{Monroe95}.
The same transformation $U_{ij}$ applied to the logical states of $\tilde{L}$, also implements a 
controlled-not gate,
\begin{eqnarray}
U_{ij}|\tilde{0}\rangle_i |\tilde{0}\rangle_j &=&|\tilde{0}\rangle_i |\tilde{0}\rangle_j \nonumber \\
U_{ij}|\tilde{0}\rangle_i |\tilde{1}\rangle_j &=&|\tilde{1}\rangle_i |\tilde{1}\rangle_j \nonumber \\
U_{ij}|\tilde{1}\rangle_i |\tilde{0}\rangle_j &=&|\tilde{1}\rangle_i |\tilde{0}\rangle_j \nonumber \\
U_{ij}|\tilde{1}\rangle_i |\tilde{1}\rangle_j &=&|\tilde{0}\rangle_i |\tilde{1}\rangle_j
\hspace{2.5 mm},
\end{eqnarray}
except that the control and target qubits are interchanged.

Since the necessary one-bit and two-bit quantum gates can be realized using known unitary 
transformations on the ion register, any arbitrary quantum computation can be implemented in 
the logical basis $\tilde{L}$.
Given an arbitrary state $|\psi\rangle_a=c_0|0\rangle_a-c_1|1\rangle_a$ for ion $a$, the 
corresponding qubit $|\tilde{\psi}\rangle_a$ in the $\tilde{L}$ basis,
\begin{equation}
|\tilde{\psi}\rangle_a=c_0|\tilde{0}\rangle_a+c_1|\tilde{1}\rangle_a
\hspace{2.5 mm},
\end{equation}
can be prepared by applying a $V_a^{\pi/2}(-\pi/2)$ laser pulse, as shown in 
Eqs.(\ref{pi/2-pulse1}-\ref{pi/2-pulse2}).

The encoding process of our error correction scheme is implemented by entangling this arbitrary 
qubit with a second qubit from another ion $b$.
Assuming that $b$ is in the ground state $|0\rangle_b$ initially, a $V_b^{\pi/2}(-\pi/2)$ 
laser pulse is applied to prepare it in the logical state $|\tilde{0}\rangle_b$.
Then a controlled-not gate $U_{ba}$ is applied to the ions (using $a$ as the control qubit) to 
obtain the desired codeword,
\begin{equation}
|\tilde{\psi}\rangle_{i}=
c_0|\tilde{0}\rangle_a\otimes|\tilde{0}\rangle_b +
c_1|\tilde{1}\rangle_a\otimes|\tilde{1}\rangle_b
\hspace{2.5 mm},
\end{equation}
which can be maintained indefinitely using coherent feedback immediately following the 
detection of spontaneous emission events.
When required, the initial qubit $|\tilde{\psi}\rangle_a$ can be recovered from this codeword 
by applying another controlled-not gate $U_{ba}$ to disentangle the ions.

If spontaneous emission occurs from the excited level of ion $a$, so that the state of the system 
is reduced to 
\begin{equation}
c_0|0\rangle_a\otimes|\tilde{0}\rangle_b-
c_1|0\rangle_a\otimes|\tilde{1}\rangle_b
\hspace{2.5 mm},
\end{equation}
the amplitude coefficients of the original qubit are preserved, and the codeword 
$|\tilde{\psi}\rangle_{i}$ can be restored by the coherent feedback process presented below,
{ which is also illustrated in Figure~\ref{feedback_1}.}

First a $V_a^{\pi/2}(\pi/2)$ laser pulse is applied to rotate the state of ion $a$ into a 
logical state, i.e., $|0\rangle_a \rightarrow |\tilde{1}\rangle_a$,
\begin{equation}
c_0|\tilde{1}\rangle_a\otimes|\tilde{0}\rangle_b-c_1|\tilde{1}\rangle_a\otimes|\tilde{1}\rangle_b
\hspace{2.5 mm},
\end{equation}
then a controlled-not gate $U_{ab}$ is applied to entangle the ions,
\begin{equation}
c_0|\tilde{1}\rangle_a\otimes|\tilde{0}\rangle_b-c_1|\tilde{0}\rangle_a\otimes|\tilde{1}\rangle_b
\hspace{2.5 mm},
\end{equation}
and finally a $V_a^{\pi}(-\pi/2)$ laser pulse is applied to flip the logical states of qubit $a$,
\begin{eqnarray}
|\tilde{0}\rangle_a &\rightarrow& -|\tilde{1}\rangle_a \\
|\tilde{1}\rangle_a &\rightarrow&  |\tilde{0}\rangle_a
\hspace{2.5mm},
\end{eqnarray}
thus restoring the codeword $|\tilde{\psi}\rangle_{i}$.

We have assumed that we know which ion has emitted spontaneously, so that we can apply this 
feedback process selectively to the appropriate ions.
If spontaneous emission occurs from the excited level of ion $b$, the same process is used to 
recover the codeword $|\tilde{\psi}\rangle_{i}$, with the ions $a,b$ interchanged.



{ To achieve codeword invariance under the conditional time evolution 
given in Eq.(\ref{cond_evolution}), the number of excited 
states $|1\rangle_j$ has to be constant for each elementary state in basis 
${\cal L}$ making up a codeword, so that the damping terms due to the 
conditional time evolution factor out without distorting the codeword.
Thus the proposed codeword $|\tilde{\psi}\rangle_i$ can be preserved between 
quantum jumps by a process of complementary extension, which results in a 
doubling of the register size.

The elementary codewords $|\tilde{0}\rangle_a|\tilde{0}\rangle_b$, 
$|\tilde{1}\rangle_a|\tilde{1}\rangle_b$ can be made invariant with respect to the conditional 
time evolution by entangling each ${\cal L}$-basis product state of the ions $a,\,b$ with its 
complementary product state from a second set of ions $c,\,d$, resulting in a 
symmetrization of the codewords.
This encoding process can be achieved by setting the ions $c,\,d$ in the excited state 
$|1\rangle_c|1\rangle_d$ initially and applying two controlled-not gates 
$U_{ac}U_{bd}$ to obtain the extended codeword $|\Psi\rangle_i$, which 
can be written in terms of the electronic states for basis ${\cal L}$ as
\begin{eqnarray}
|\Psi\rangle_i &=& c_0\left(|00\rangle_1|11\rangle_2+|01\rangle_1|10\rangle_2+
|10\rangle_1|01\rangle_2+|11\rangle_1|00\rangle_2\right)+ \nonumber\\
&& \!\!\!\!\!\mbox{}+c_1\left(|00\rangle_1|11\rangle_2-|01\rangle_1|10\rangle_2-
|10\rangle_1|01\rangle_2+|11\rangle_1|00\rangle_2\right)
\hspace{2.5 mm},
\end{eqnarray}
where $|S_aS_b\rangle_1|S_cS_d\rangle_2=
|S_a\rangle_a\otimes|S_b\rangle_b\otimes|S_c\rangle_c\otimes|S_d\rangle_d$, and $S_i=0,1$.

However a new feedback process is now required to invert the effect of quantum jumps on this 
extended codeword.
First a $V^\pi(-\pi/2)$ laser pulse is applied to the ion from which the decay event is 
detected, and then the product states of all four ions are complemented to recover 
$|\Psi\rangle_i$, using the transformation
\begin{equation}
|S_aS_b\rangle_1|S_cS_d\rangle_2 \rightarrow
\frac{1}{\sqrt{2}} \left(|S_aS_b\rangle_1|S_cS_d\rangle_2+
|\bar{S}_a\bar{S_b}\rangle_1|\bar{S_c}\bar{S_d}\rangle_2\right)
\hspace{2.5 mm},
\end{equation}
where $\bar{S_i}$ is the complement of $S_i$ (modulo $2$).
This feedback process has been adopted in the following section to invert the effect of 
quantum jumps on codewords consisting of complementary electronic number states, in order 
to implement an improved error-correcting code, which requires less ancillary qubits.}


\section{Error correction using electronic number states}

The electronic number states $|k\rangle_e$ of a system of N trapped ions form a logical basis, 
and are defined by the product of the individual electronic levels of the ions arranged in 
some definite order,
\begin{equation}
|k\rangle_e=|S_{N}\rangle_{N}\otimes|S_{N-1}\rangle_{N-1}\otimes\cdots\otimes|S_{1}\rangle_{1}
\hspace{2.5 mm},
\end{equation}
where $S_i=0,1$ represents a ground or excited state respectively.
An N-ion register has $2^N$ electronic number states labeled by the integer 
$k=S_{N}\times2^{N-1}+S_{N-1}\times2^{N-2}+\cdots+S_{1}\times2^{0}$ $(0 \leq k \leq 2^N-1)$ 
representing the binary string $S_{N}S_{N-1}\dots S_1$.

Spontaneous emission from the excited level $|1\rangle_j$ of the $j$-th ion eliminates the 
amplitude coefficients of the number states $|q\rangle_e$ which include the ground state 
$|0\rangle_j$, and are labeled by the integers $q$ containing $S_j=0$.
However the amplitude coefficients of the complementary number states $|\bar{q}\rangle_e$ are 
preserved, where $\bar{q}$ is the complement of $q$ (modulo $2^N$).
Therefore if the initial state $|\psi\rangle_i$ is prepared so that the amplitude coefficients 
of the electronic number states $|k\rangle_e$ and $|\bar{k}\rangle_e$ are equal for all $k$, 
\begin{equation}
|\psi\rangle_i=\sum_{k=0}^{2^{N-1}-1}\frac{c_k}{\sqrt{2}}(|k\rangle_e+|\bar{k}\rangle_e)
\hspace{2.5 mm},
\end{equation}
then $|\psi\rangle_i$ can be restored after a spontaneous emission event from any of the ions.

Given an arbitrary (N-1)-ion state 
\begin{equation}
|\psi\rangle_i^{(N-1)}=\sum_{k=0}^{2^{N-1}-1}c_k|k\rangle_e^{(N-1)}
\hspace{2.5 mm},
\end{equation}
the required N-ion codeword $|\psi\rangle_i$ can be generated by coupling $|\psi\rangle_i^{(N-1)}$ 
to the ground state of an additional ion 
\begin{equation}
\sum_{k=0}^{2^{N-1}-1}c_k|0\rangle_N\otimes|k\rangle_e^{(N-1)}
\hspace{2.5 mm},
\end{equation}
and complementing each N-ion electronic number state 
$|k\rangle_e=|0\rangle_N\otimes|k\rangle_e^{(N-1)}$.
The original (N-1)-ion state $|\psi\rangle_i^{(N-1)}$ can be recovered from the codeword 
by applying this transformation in reverse to disentangle the N-th ion from the other qubits.
The complementing procedure is also used to restore the codeword after a spontaneous emission 
event, 
{ as shown in Figure~\ref{feedback_2}, }
and its implementation is presented below.

We consider a three ion register to demonstrate our error-correcting code.
An arbitrary 2-ion ($a, b$) state given by
\begin{equation}
c_0|0\rangle_e+c_1|1\rangle_e+c_2|2\rangle_e+c_3|3\rangle_e
\hspace{2.5 mm},
\end{equation}
is transformed into the required codeword $|\psi\rangle_i$ by coupling it to a third ion $c$ and 
applying a complementing procedure,
\begin{equation}
|\psi\rangle_i = \frac{1}{\sqrt{2}} (
c_0|0\rangle_e+c_1|1\rangle_e+c_2|2\rangle_e+c_3|3\rangle_e +
c_3|4\rangle_e+c_2|5\rangle_e+c_1|6\rangle_e+c_0|7\rangle_e )
\hspace{2.5 mm},
\end{equation}
where the electronic number states 
$|k\rangle_e=|S_{c}\rangle_{c}\otimes|S_{b}\rangle_{b}\otimes|S_{a}\rangle_{a}$ 
for the ions $a,b,c$, are labeled by $k=4S_c+2S_b+S_a$.

Hence, if spontaneous emission occurs from the excited level of ion $b$ for example, the codeword 
is reduced to 
\begin{equation}
c_2|0\rangle_e+c_3|1\rangle_e+c_1|4\rangle_e+c_0|5\rangle_e
\hspace{2.5 mm},
\end{equation}
and the original amplitude coefficients are preserved.

The feedback process necessary to restore the codeword $|\psi\rangle_i$ consists of two steps.
First a $V_b^\pi(-\pi/2)$ laser pulse is applied to ion $b$, i.e., 
$|0\rangle_b \rightarrow |1\rangle_b$, 
to match the amplitude coefficients with their initial electronic number states, 
\begin{equation}
c_2|2\rangle_e+c_3|3\rangle_e+c_1|6\rangle_e+c_0|7\rangle_e
\hspace{2.5 mm},
\end{equation}
and then each electronic number state is complemented, ie.\ 
$|k\rangle_e \rightarrow \frac{1}{\sqrt{2}}(|k\rangle_e+|\bar{k}\rangle_e)$, 
to obtain the codeword $|\psi\rangle_i$.

{ The required complementing transformation ${\cal C}_N$ }
is an extension of 
the two-bit controlled-not gate to N+1 ions, and therefore requires a 
vibrational mode common to all the ions.
An extra ion $x$ is prepared in the superposition $\frac{1}{\sqrt{2}}(|0\rangle_x+|1\rangle_x)$, 
and then entangled with the other ions.
Each electronic number state $|k\rangle_e$ coupled to ion $x$, 
\begin{equation}
\frac{1}{\sqrt{2}}(|0\rangle_x+|1\rangle_x)\otimes(|S_{N}\rangle_{N}...|S_{1}\rangle_{1})
\label{xsuper}
\hspace{2.5 mm},
\end{equation}
is transformed by a series of controlled-not gates $U_{x1}U_{x2}...U_{xN}$ (applied sequentially 
in any order), which leave qubit $x$ unchanged,
\begin{eqnarray}
|0\rangle_x|S_{j}\rangle_{j} & \rightarrow & |0\rangle_x|S_{j}\rangle_{j} \\
|1\rangle_x|S_{j}\rangle_{j} & \rightarrow & |1\rangle_x|\bar{S}_{j}\rangle_{j}
\hspace{2.5 mm},
\end{eqnarray}
while complementing $|k\rangle_e$, to transform the state from Eq.(\ref{xsuper}) into
\begin{equation}
\frac{1}{\sqrt{2}}(|0\rangle_x\otimes
|S_{N}\rangle_{N}...|S_{1}\rangle_{1}+
|1\rangle_x\otimes
|\bar{S}_{N}\rangle_{N}...|\bar{S}_{1}\rangle_{1})
\hspace{2.5 mm},
\end{equation}
where $\bar{S}_{j}$ is the complement of $S_j$ (modulo 2).
We note that the controlled-not gates in the sequence $U_{x1}U_{x2}...U_{xN}$ could be 
implemented simultaneously, using an (N+1)-bit quantum gate.

In the feedback process used to restore $|\psi\rangle_i$, a $V_j^\pi(-\pi/2)$ laser pulse 
is first applied after spontaneous emission from the $j$-th ion, which sets the $j$-th 
digit of the remaining electronic number states to $S_j=1$.
Hence after applying the controlled-not sequence $U_{x1}U_{x2}...U_{xN}$, 
the ancillary ion $x$ can be disentangled from the $N$-ion number states by applying 
one more controlled-not gate $U_{jx}$, using ion $j$ which decayed spontaneously as 
the control qubit.
The resultant state of the ion register is given by
\begin{equation}
|1\rangle_x\otimes\frac{1}{\sqrt{2}}(
|S_{N}\rangle_{N}...|S_{1}\rangle_{1}+
|\bar{S}_{N}\rangle_{N}...|\bar{S}_{1}\rangle_{1})
\hspace{2.5 mm},
\end{equation}
for the electronic number state $|k\rangle_N$, thus realizing the required complementing 
transformation
\begin{equation}
{\cal C}_N|k\rangle_e = \frac{1}{\sqrt{2}}(|k\rangle_e+|\bar{k}\rangle_e)
\hspace{2.5 mm}.
\end{equation}

In the case that the complementing transformation is used to encode the arbitrary state 
$|\psi\rangle_i^{(N-1)}$, 
the $N$-th digit of the $N$-ion number states has been set to $S_N=0$ by preparing the 
$N$-th ion in the ground state, therefore the last step disentangling the ancillary ion should 
consist of applying the controlled-not gate $U_{Nx}$, using ion $N$ as the control qubit.

Once again we have assumed that the continuous monitoring of spontaneous emission enables 
us to know which ion has emitted spontaneously, so that the coherent feedback process can be 
applied selectively.
The resonant $\pi$-pulse $V^\pi(-\pi/2)$ is applied to the particular ion from which spontaneous 
emission is detected, and is followed by the complementing procedure to recover the initial 
codeword.



{ As in the previous section, the codeword $|\psi\rangle_i$ can be made 
invariant under the conditional time evolution between quantum jumps by 
entangling each electronic number state with its complementary 
number state from a second set of $N$ ions.
Thus the number of excited states $|1\rangle_j$ is constant for each elementary $2N$-ion 
state making up the extended codeword, and equal to $N$.
This encoding process can be achieved for $|\psi\rangle_i$ by setting the second set 
of ions in the state $|2^N-1\rangle_2$ initially and applying a series of $N$ 
controlled-not gates $U_{a_1a_2}U_{b_1b_2}\ldots$ to pairs of ions 
from the first $(1)$ and second $(2)$ sets, in order to obtain the symmetrized codeword 
$|\Psi\rangle_{2N}$,
\begin{equation}
|\Psi\rangle_{2N}=\sum_{k=0}^{2^{N-1}-1}\frac{c_k}{\sqrt{2}}
\left(|k\rangle_1|\bar{k}\rangle_2+|\bar{k}\rangle_1|k\rangle_2\right)
\hspace{2.5 mm}.
\end{equation}

In this case, the same feedback process presented above can be used to 
invert the effect of quantum jumps on the codeword $|\Psi\rangle_{2N}$, 
using the complementing transformation
\begin{equation}
{\cal C}_{2N}|k\rangle_1|\bar{k}\rangle_2 =
\frac{1}{\sqrt{2}}\left(|k\rangle_1|\bar{k}\rangle_2+
|\bar{k}\rangle_1|k\rangle_2\right)
\hspace{2.5 mm},
\end{equation}
which involves the electronic number states of all $2N$ ions in the register.}


\section{Discussion and Conclusion}


{ 
The efficiency of the proposed error-correcting codes can be compared by considering their 
demands on memory and time resources.
In both cases the requirement for ancillary qubits is reduced compared to the general 
schemes discussed in Section \ref{Introduction}, which require the codeword 
encoding each logical qubit to use at least five qubits for ``perfect'' error correction, 
as shown by Laflamme et al.\ \cite{Laflamme96}, and cannot correct for the conditional 
time evolution.

The quantum error correction scheme based on Fourier-transformed states can be extended 
to more logical qubits in a straightforward manner, eliminating the effects of spontaneous 
emission using four-qubit codewords, and thus storing $N$ logical qubits on a $(4N+1)$-ion 
register, taking into account the extra ion required for the complementing transformation.

On the other hand, the error-correcting code employing $2N$-ion complementary number states 
can store $N$ logical qubits on a $(2(N+1)+1)$-ion register, and evidently uses available 
memory resources (trapped ions) more efficiently, especially for large numbers of logical 
qubits.
However the price paid for this saving is that the feedback process involves acting on all 
the ions to correct a spontaneous emission event, rather than only four ions, hence this code 
requires more time (logic gate operations) and is harder to implement experimentally.
}

The time required for the feedback process can be estimated from the number of quantum 
gates used in its implementation, which is shown in Table \ref{table} for 
{ the storage of $N$ logical qubits, using both error correction schemes.}

The controlled-not gate realized by Monroe {\em et al.}\ \cite{Monroe95} in a Paul trap, has a 
duration of $\tau_{cn}\approx 50\mu$s, however it only operates on the qubits of a single ion 
and the associated vibrational mode.
This accounts for approximately half of the laser pulses required to operate on two ions, 
since the state of the vibrational mode is not transferred to the second ion.
The switching rate of the laser pulses implementing this logic gate is bounded by the 
characteristic vibrational frequency in the Paul trap ($\nu\leq$100MHz).
We expect that in the first experimental realizations using a linear rf trap, the duration of a 
controlled-not gate between two ions will be greater than that obtained in a Paul trap, since 
the vibrational frequency will be of the order of only 10-100 kHz, however a value of 
$\tau_{cn}\approx 100\mu$s seems feasible eventually.

The time taken to realize a qubit rotation is inversely proportional to the Rabi frequency of the 
applied laser pulse.
In the same article referred to above, Monroe {\em et al.}\ report that the Rabi frequency of the 
carrier pulses was 140kHz, which results in interaction times of approximately $10\mu$s and 
$20\mu$s, for $\pi/2$ and $\pi$-pulses respectively.
The interaction times necessary to implement both qubit rotations and controlled-not gates, are 
expected to be reduced in future experiments as the laser power available is increased.
However the maximum laser intensity is limited because at high intensities the two-level 
approximation and the rotating wave approximations will break down.


{ 
The feedback time for the error-correcting code using complementary electronic number states 
is proportional to the number of ions used, and is approximately given by $(2N+1)\tau_{cn}$.
Therefore the time required to implement 5 logical qubits (using a 13-ion register) is 
$\approx 1.1$s, which is still short compared to the decoherence time 
$\tau_d\approx5$s of the register (assuming a qubit decoherence time $\tau_q\approx1$ 
minute), and ensures that spontaneous emission is unlikely to occur before the codeword 
is restored.

In comparison, the error-correcting code using Fourier-transformed states only requires 
a constant time $\approx0.5$s for its feedback process, though it requires a 21-ion 
register to implement 5 logical qubits.
}

There are a number of problems which can affect the performance of our schemes including 
inefficient detection of spontaneous emission events, gate errors during the feedback process and 
decoherence due to vibrational damping while the ions are stored.
We have assumed that these problems will be manageable in future experiments by some combination 
of technical accuracy and error-correcting codes, in order to focus on how the errors due to a 
single fundamental source of decoherence can be corrected.

For example, the continuous monitoring of spontaneous emission over a $4\pi$ solid angle cannot 
be realized efficiently with present technology since the photodetection efficiency is very small 
($<10^{-2}$), and it is very difficult to cover the entire angle with detectors.
The implementation of two-qubit logic gates in the feedback, encoding and decoding processes 
was also assumed to be perfect, but could be realized by making use of the fault-tolerant 
controlled-not gates proposed by Cirac, Pellizzari and Zoller \cite{Cirac96}.

In conclusion, we have proposed two quantum error correction schemes which rely on continuous 
monitoring of spontaneous emission and selective coherent feedback to eliminate memory errors as 
soon as they are detected.
A minimum number of ancillary qubits are required to realize codewords, by using a Fourier-transformed 
logical basis for the first scheme, and complementary electronic number states for the second scheme.

Acknowledgements:- C. D'Helon would like to thank H. Mabuchi for his comments 
and suggestions, 
{ and M. Plenio for pointing out his article `Quantum error 
correction in the presence of spontaneous emission' 
(Reference \cite{Plenio97}).}

\newpage

\begin{table}
\vspace{10mm}
\begin{tabular}{|c||c|c|} \hline
error correction scheme & number of & number of\\
using&qubit rotations&controlled-not gates\\ \hline
Fourier-transformed states&2&5\\ \hline
Complementary number states&2&2N+1\\ \hline
\end{tabular}
\vspace{10mm}
\caption{
{ 
Number of elementary logic gates required to implement the error-correcting schemes for $N$ logical qubits.
}
}
\label{table}
\end{table}

\newpage

\begin{figure}
\caption{ Flow chart of the encoding process for a codeword $|\tilde{\psi}\rangle_i$, 
and the required feedback process if a spontaneous decay event occurs from ion $a$.}
\label{feedback_1}
\end{figure}

\begin{figure}[tb]
\caption{ Flow chart of the encoding process for a codeword $|\psi\rangle_i$, and the 
required feedback process if the spontaneous decay event occurs from ion $j$, where the 
transformation ${\cal A}_N$ represents the series of controlled-not gates 
$U_{x1}U_{x2}...U_{xN}$.}
\label{feedback_2}
\end{figure}

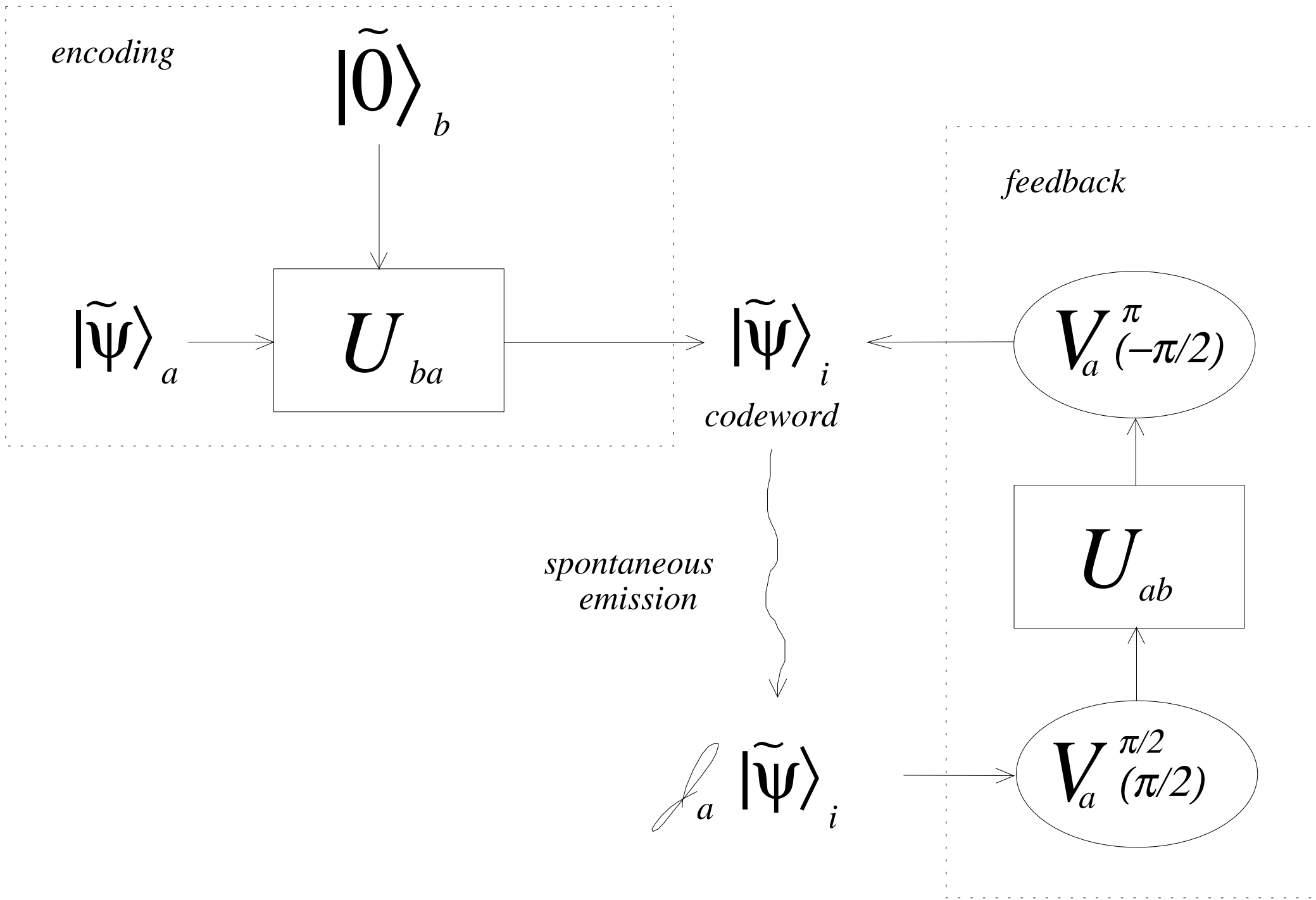
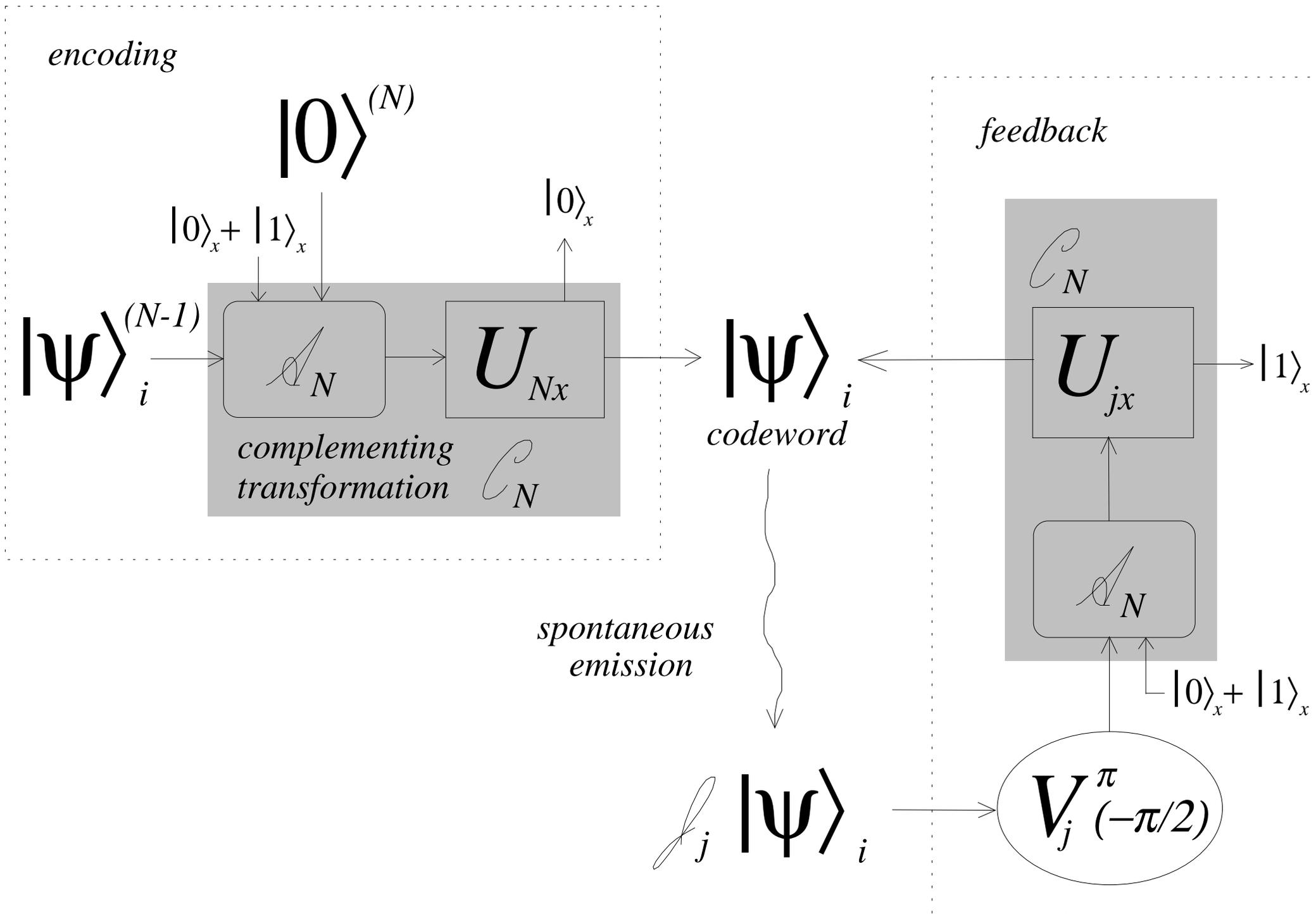
\end{document}